# General Conversion between ANCF and B-spline Surfaces

Randi Wang[a,b], Peng Lan[a,✉], Zuqing Yu[a], Nianli Lu[a]

a. School of Mechatronic Engineering, Harbin Institute of Technology, Harbin, 150001, China; b. Department of Mechanical Engineering, University of Wisconsin-Madison, Madison, 53706, WI, USA

✉ Email: lan_p@sina.com (Peng Lan)

**Abstract:** In this paper, general conversion equations are derived between Absolute Nodal Coordinates Formulation (ANCF) finite surface elements and $k \times l \, (k, l \leq 3, k, l \in N^+)$ B-spline surfaces, an extension of our previous work on the conversion between ANCF cable elements and B-spline curves.

The derivation of the conversion equations is the discovery of the geometric invariance of ANCF displacement field before and after the conversion. Our study starts from proposing the conversion equation between ANCF finite surface elements and $3 \times 3$ Bezier surfaces which are the special cases of B-spline surfaces, followed by establishing a general conversion equation between ANCF finite surface elements and $m \times n \, (m, n \leq 3, m, n \in N^+)$ Bezier surfaces. This general conversion equation has functionalities (1) to realize the one-step direct conversion between ANCF and $m \times n \, (m, n \leq 3, m, n \in N^+)$ Bezier surfaces (2) to convert ANCF finite surface elements directly to $m \times n \, (m, n < 3, m, n \in N^+)$ Bezier surfaces provided the ANCF nodal coordinates are not independent. The direct conversion from a conditional ANCF finite surface to a $m \times n \, (m, n < 3, m, n \in N^+)$ Bezier surfaces enhances the efficiency and ability to control and store data in computers during the conversion process.

The conversion between ANCF finite surface elements and B-spline surfaces is derived from a conversion of $3 \times 3$ B-spline surfaces to a more general conversion of $k \times l \, (k, l \leq 3, k, l \in N^+)$ B-spline surfaces. B-spline basis functions are utilized in the non-recursive form, from which a more efficient conversion equation is obtained compared with an intuitive conversion semantics where one converts firstly B-spline surfaces to composite Bezier surfaces by inserting knot and converts to ANCF finite surface elements afterwards. The obtained conversion equations between ANCF and B-spline surfaces realize the one-step direct conversion. Conversion equations between ANCF and Bezier, B-spline surfaces proposed are a further exploration on the integration between computer aided design (CAD) and computer aided analysis (CAA) based on integration of ANCF finite elements and NURBS geometry.

# 1. Introduction

In a product life cycle, the geometric design iteratively subjects to engineering analysis, followed by design changes or optimization. This iterative product development process decreases dramatically the efficiency of production. As society becomes more accustomed to the high efficient innovative and improved technology, there is a requirement for the product development process to speed up. The solution is the integration of computer aided design (CAD) and computer aided analysis (CAA) [1-3]. Motivation of the integration of CAD and CAA requires a seamless interface between geometric design and engineering analysis. The essential issue on this integration comes from the incompatible geometric representations[4,5]. Models in CAD systems are combinations of high-level parametric representations and evaluated boundary conditions[6,7]. One typical approach to represent geometry in CAD systems is Bezier, B-spline and non-uniform rational B-spline (NURBS) representations. Models in CAA systems are combinations of boundary conditions, constitutive equations and governing equations. These CAA models are engineering analysis problems to be solved in terms of 3 above conditions in the domain obtained from corresponding CAD models. However, common polynomial interpolations of the traditional finite element method (FEM), a typical approach of CAA, may not exactly describe the obtained domain. The incompatibility of CAD and CAA is caused by this inconsistent geometric description[8-10]. The conversion between B-splines curves and ANCF cable elements is a successful integration between CAD and CAA where the conversion process is completely automated in terms of the established transformation matrix[1,11]. Compared to traditional FEM upon structural elements description, ANCF elements introduce slope vectors instead of infinite rotation as nodal variables and these nodal variables are described in an absolute inertial coordinate. ANCF grows a wiser approach for analysis of large multi-body deformation system[12-18]. It is proved the geometry has the same geometry expressed in terms of ANCF cable elements and converted ANCF cable elements whose nodal coordinates are represented by control points of B-splines curves[1,11,12]. This makes it possible that the geometric description of the geometric domain in CAD systems and the solution domain of engineering analysis in CAA systems are compatible thus real-time modification becomes possible when changes are made in either geometric design or engineering analysis.

The contribution of this paper is an extension of 1-dimensional study to derive the conversion equations between ANCF finite surface elements and 2-dimensional Bezier and B-spline surfaces. The derivation of these conversion equations is based on the discovery of the geometric invariance of ANCF displacement field before and after the conversion. Our study finds:

(1) Approach to represent 16 ANCF absolute nodal coordinates by 16 control points of $3 \times 3$ Bezier surfaces, followed by establishing the conversion equations between ANCF finite surface elements and $3 \times 3$ Bezier surfaces.
(2) General transformation matrix and general conversion equations between ANCF surface elements and $m \times n \, (m, n \leq 3, m, n \in N^+)$ Beziers surfaces.
(3) Independence of position and gradient variables affects the degrees of converted surfaces and finite surfaces element hence the efficiency of the conversion process.
(4) General transformation matrix and general conversion equations between ANCF

surface elements and $k \times l$ $(k,l \leq 3, k,l \in N^+)$ B-spline surfaces.

In the remainder of this paper, Section 2 illustrates the conversion between the $3 \times 3$ Bezier surfaces and ANCF finite surface elements. Section 3 proposes the conversion between ANCF finite surface elements and $m \times n$ ($m,n \leq 3$, $m,n \in N^+$) Beziers surfaces. The independence of position and gradient variables is discussed. Section 4 presents the conversion equations between ANCF finite surface elements and B-spline surfaces. This study is concluded in Section 5.

## 2. Conversion between Bezier surfaces and ANCF finite surface elements

### 2.1 Bezier surfaces representation

This section gives a brief introduction of Bezier surfaces. Bezier surfaces are commonly utilized to geometric representation in CAD systems. Bezier, B-splines and non-uniform rational B-splines have become the standard geometric representations in CAD systems. The introduction of fundamental properties of Bezier surfaces is useful for illustrating the integration between CAD and CAA. Details on Bezier representation refer to *The NURBS Book* [19].

A Bezier function of degrees $m \times n$ is defined as follows

$$\mathbf{p}(u,v) = \sum_{i=0}^{m} \sum_{j=0}^{n} \mathbf{b}_{ij} B_{i,m}(u) B_{j,n}(v), 0 < u,v < 1 \tag{1}$$

where, $\mathbf{b}_{ij}(i=0,...,m; j=0,...,n)$ represent a set of Bezier coefficients called control points. $B_{i,m}(u)$ and $B_{j,n}(v)$ are Bernstein polynomials defined in the square domain $(u,v) \in [0,1] \times [0,1]$. The function $\mathbf{p}(u,v)$ maps the unit square domain into a smooth-continuous image embedded within a space of the same dimensionality as $\mathbf{b}_{ij}$.

Eq. (2) gives an intuitive Bezier surface representation in terms of geometric matrix and Hermite matrix, in which $\mathbf{b}_{ij}$ is the geometric matrix. $\mathbf{B}_{i,m}(u)$ and $\mathbf{B}_{j,n}(v)$ are obtained from the Hermite matrix $\mathbf{M}_m$ and $\mathbf{M}_n$ in Eq.(4).

$$\mathbf{p}(u,v) = \mathbf{B}_{i,m}(u) \cdot \mathbf{b}_{ij} \cdot \mathbf{B}_{j,n}(v) \tag{2}$$

where,

$$\mathbf{b}_{ij} = \begin{bmatrix} \mathbf{b}_{00} & \mathbf{b}_{01} & \cdots & \mathbf{b}_{0n} \\ \mathbf{b}_{10} & \mathbf{b}_{11} & \cdots & \mathbf{b}_{1n} \\ \vdots & \vdots & & \vdots \\ \mathbf{b}_{m0} & \mathbf{b}_{m1} & \cdots & \mathbf{b}_{mn} \end{bmatrix}$$

$$\mathbf{B}_{i,m}(u) = [B_{0,m}(u) \ B_{1,m}(u) \ ... \ B_{m,m}(u)] = \mathbf{u} \cdot \mathbf{M}_m \tag{3}$$

$$\mathbf{B}_{j,n}(v) = \begin{bmatrix} B_{0,n}(v) & B_{1,n}(v) & \cdots & B_{n,n}(v) \end{bmatrix}^T = \mathbf{v} \cdot \mathbf{M}_n$$

In Eq.(3), $\mathbf{u} = [1 \ u \ ... \ u^m]$ and $\mathbf{v} = [1 \ v \ ... \ v^n]$ are power basis vectors. the Hermite matrix $\mathbf{M}_m$ and $\mathbf{M}_n$ are defined as follows

$$\mathbf{M}_m = \begin{bmatrix} m_{0,0} & m_{0,1} & \cdots & m_{0,m} \\ m_{1,0} & m_{1,1} & \cdots & m_{1,m} \\ \vdots & \vdots & & \vdots \\ m_{m,0} & m_{m,1} & \cdots & m_{m,m} \end{bmatrix}, \quad \mathbf{M}_n = \begin{bmatrix} m_{0,0} & m_{0,1} & \cdots & m_{0,n} \\ m_{1,0} & m_{1,1} & \cdots & m_{1,n} \\ \vdots & \vdots & & \vdots \\ m_{n,0} & m_{n,1} & \cdots & m_{n,n} \end{bmatrix} \quad (4)$$

where, $\mathbf{M}_m$ and $\mathbf{M}_n$ are lower triangular matrices, in which elements have the form:

$$m_{ij} = (-1)^{i+j} C_r^i C_i^j \quad r = m, n, \quad C_i^j = \frac{i!}{j!(i-j)!} \quad (5)$$

Figure 1 shows a $3 \times 3$ Bezier surface over the domain $(u, v) \in [0,1] \times [0,1]$ and the corresponding 16 control points $b_{ij}(i, j = 0, 1, 2, 3)$.

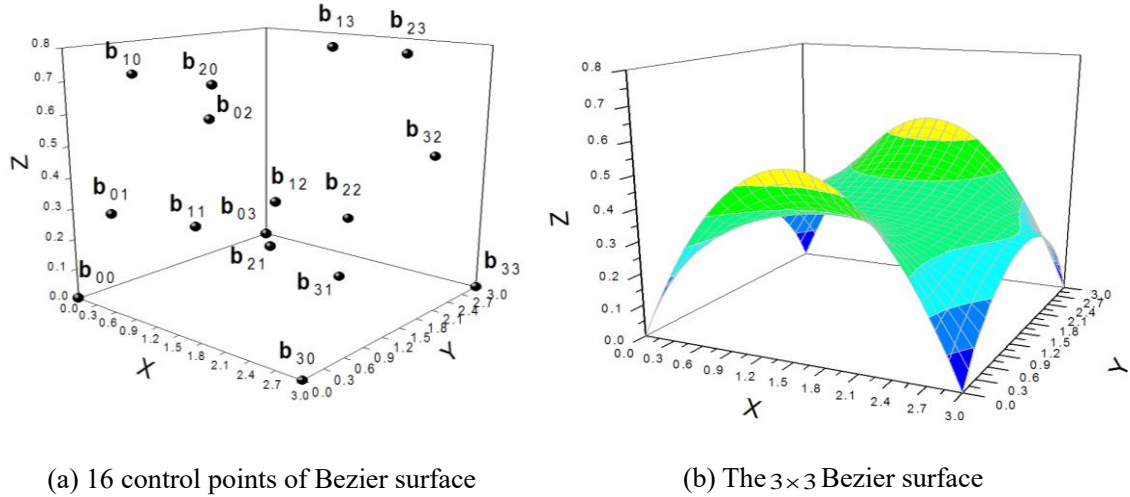

(a) 16 control points of Bezier surface  (b) The $3 \times 3$ Bezier surface

Figure 1  $p(u,v)$ maps the unit square domain into a smooth-continuous $3 \times 3$ Bezier surface embedded within a space of the same dimensionality as $b_{ij}$.

**2.2 ANCF Surface Finite Element**

This section gives a brief description of the ANCF finite surface element. It is a Hermitian element, being utilized to describe a parameterized surface, extended from a 16 Degrees Of Freedom (DOF) Hermitian rectangle element by substituting the displacement field by three separate scalars. ANCF finite surface element introduces four nodal coordinates, in which one to describe absolute nodal position coordinates, two to describe one-order position coordinates gradient, the rest one to describe second-order position coordinates gradient, 48 DOFs in total[20]. The displacement field expressed in terms of these nodal coordinates and shape functions is

$$\mathbf{r}(u,v) = \mathbf{S}\mathbf{e} \quad (6)$$

where $\mathbf{S}$ is a matrix of global shape functions; $\mathbf{e}$ is a vector of nodal coordinates. The elements in $\mathbf{S}$ are Hermitian functions composed of two-dimensional bean functions and the elements in $\mathbf{e}$ are 16 nodal coordinates. They are specified in Eq.(7)~Eq.(11).

$$\mathbf{S} = [S_{11}\mathbf{I}, S_{12}\mathbf{I}, S_{13}\mathbf{I}, S_{14}\mathbf{I}; \cdots; S_{41}\mathbf{I}, S_{42}\mathbf{I}, S_{43}\mathbf{I}, S_{44}\mathbf{I}] \tag{7}$$

where, $\mathbf{I}$ is the identity matrix and $S_{ij}$ $(i, j = 1,...,4)$ are defined as

$$S_{ij} = s_i(u,a) \cdot s_j(v,b) \tag{8}$$

in which, $a$ and $b$ stand for length and width of the ANCF finite surface element. Beam functions $s_i(u,a)$ and $s_j(v,b)$ are defined as follows

$$\begin{cases} s_1(\omega,l) = 1 - 3\lambda^2 + 2\lambda^3 \\ s_2(\omega,l) = l(\lambda - 2\lambda^2 + \lambda^3) \\ s_3(\omega,l) = 3\lambda^2 - 2\lambda^3 \\ s_4(\omega,l) = l(\lambda^3 - \lambda^2) \end{cases} \tag{9}$$

in which, $\lambda = \omega/l$ where $\omega = u, v$ and $l = a, b$.

The vector of the nodal coordinates $\mathbf{e}$ is

$$\mathbf{e} = \left[\mathbf{r}_{00}^{00}, \mathbf{r}_{00}^{10}, \mathbf{r}_{a0}^{00}, \mathbf{r}_{a0}^{10}, \mathbf{r}_{00}^{01}, \mathbf{r}_{00}^{11}, \mathbf{r}_{a0}^{01}, \mathbf{r}_{a0}^{11}, \mathbf{r}_{0b}^{00}, \mathbf{r}_{0b}^{10}, \mathbf{r}_{ab}^{00}, \mathbf{r}_{ab}^{10}, \mathbf{r}_{0b}^{01}, \mathbf{r}_{0b}^{11}, \mathbf{r}_{ab}^{01}, \mathbf{r}_{ab}^{11}\right]^T \tag{10}$$

where,

$$\mathbf{r}_{uv}^{ij} = \frac{\partial^{i+j}\mathbf{r}}{\partial \mathbf{u}^i \partial \mathbf{v}^j} \tag{11}$$

in which, $\mathbf{r}_{uv}^{ij}$ $(i = j = 0)$ represents displacement, $\mathbf{r}_{uv}^{ij}$ $(i + j = 1)$ represents one-order gradient of displacement and $\mathbf{r}_{uv}^{ij}$ $(i + j = 2)$ represents second-order gradient of displacement.

## 2.3 Conversion between ANCF and Bezier surfaces

ANCF finite surface element introduces absolute nodal position coordinates and corresponding gradients as its nodal coordinates. If the origin of the global coordinate is fixed, position vectors can commonly be represented in terms of coordinates of points and gradient vectors can always be given as the difference vectors in terms of position vectors. Therefore, the gradient vectors are associated to difference of position vectors of points. A 48 D-O-F ANCF surface element introduces 16 independent nodal coordinates, 12 of which are independent gradients of position vectors thereby 12 position vectors required to substitute them. The rest four position vectors can lie on the surface element. By comparison, 16 independent control points are required to describe a $3 \times 3$ Bezier surface. 12 of which are independent lying over the surface and the rest four of which lie at the four vertices of the surface. Thus, an ANCF finite surface element and a $3 \times 3$ Bezier surface can describe the same surface in terms of the proposed conversion equation in Eq.(12).

$$\mathbf{e} = \mathbf{T}\mathbf{b}_{ij} = \begin{bmatrix} \mathbf{A} & \mathbf{0} \\ \mathbf{0} & \mathbf{B} \end{bmatrix} \mathbf{b}_{ij} \tag{12}$$

where, $\mathbf{e}$ is the vector of 16 ANCF absolute nodal coordinates. $\mathbf{b}_{ij}$ is the vector of 16 Bezier surface control points. $\mathbf{T}$, a diagonal block-partitioned matrix, is the transformation matrix of them. Blocks $\mathbf{A}$ and $\mathbf{B}$ of $\mathbf{T}$ are lower triangular matrices. The elements of $\mathbf{A}, \mathbf{B}$ and $\mathbf{T}$ refer to Eq.(13) ~ Eq.(16).

An ANCF finite surface element and a $3 \times 3$ Bezier surface can describe the same surface in terms of the conversion equation in Eq.(12) because the two expressions are the same, one of which is the expression of the $3\times 3$ Bezier surface expressed by 16 control points shown in Eq.(14), the other is the expression obtained by substituting the very left hand side (LHS) **e** expressed by $\mathbf{b}_{ij}$ in Eq.(12) to Eq.(6).

$$\mathbf{e} = \left[\mathbf{r}_{00}^{00}, \mathbf{r}_{00}^{10}, \mathbf{r}_{a0}^{00}, \mathbf{r}_{a0}^{10}, \mathbf{r}_{00}^{01}, \mathbf{r}_{00}^{11}, \mathbf{r}_{a0}^{01}, \mathbf{r}_{a0}^{11}, \mathbf{r}_{0b}^{00}, \mathbf{r}_{0b}^{10}, \mathbf{r}_{ab}^{00}, \mathbf{r}_{ab}^{10}, \mathbf{r}_{0b}^{01}, \mathbf{r}_{0b}^{11}, \mathbf{r}_{ab}^{01}, \mathbf{r}_{ab}^{11}\right]^T \qquad (13)$$

$$\mathbf{b}_{ij} = \left[\mathbf{b}_{00}\ \mathbf{b}_{10}\ \mathbf{b}_{30}\ \mathbf{b}_{20}\ \mathbf{b}_{01}\ \mathbf{b}_{11}\ \mathbf{b}_{31}\ \mathbf{b}_{21}\ \mathbf{b}_{03}\ \mathbf{b}_{13}\ \mathbf{b}_{33}\ \mathbf{b}_{23}\ \mathbf{b}_{02}\ \mathbf{b}_{12}\right]^T \qquad (14)$$

$$\mathbf{A} = \begin{bmatrix}
\mathbf{I} & 0 & 0 & 0 & 0 & 0 & 0 & 0 \\
-\frac{3}{a}\mathbf{I} & \frac{3}{a}\mathbf{I} & 0 & 0 & 0 & 0 & 0 & 0 \\
0 & 0 & \mathbf{I} & 0 & 0 & 0 & 0 & 0 \\
0 & 0 & \frac{3}{a}\mathbf{I} & -\frac{3}{a}\mathbf{I} & 0 & 0 & 0 & 0 \\
-\frac{3}{b}\mathbf{I} & 0 & 0 & 0 & \frac{3}{b}\mathbf{I} & 0 & 0 & 0 \\
\frac{9}{ab}\mathbf{I} & -\frac{9}{ab}\mathbf{I} & 0 & 0 & -\frac{9}{ab}\mathbf{I} & \frac{9}{ab}\mathbf{I} & 0 & 0 \\
0 & 0 & -\frac{3}{b}\mathbf{I} & 0 & 0 & 0 & \frac{b}{3}\mathbf{I} & 0 \\
0 & 0 & -\frac{9}{ab}\mathbf{I} & \frac{9}{ab}\mathbf{I} & 0 & 0 & \frac{9}{ab}\mathbf{I} & -\frac{9}{ab}\mathbf{I}
\end{bmatrix} \qquad (15)$$

$$\mathbf{B} = \begin{bmatrix}
\mathbf{I} & 0 & 0 & 0 & 0 & 0 & 0 & 0 \\
-\frac{3}{a}\mathbf{I} & \frac{3}{a}\mathbf{I} & 0 & 0 & 0 & 0 & 0 & 0 \\
0 & 0 & \mathbf{I} & 0 & 0 & 0 & 0 & 0 \\
0 & 0 & \frac{3}{a}\mathbf{I} & -\frac{3}{a}\mathbf{I} & 0 & 0 & 0 & 0 \\
\frac{3}{b}\mathbf{I} & 0 & 0 & 0 & -\frac{3}{b}\mathbf{I} & 0 & 0 & 0 \\
-\frac{9}{ab}\mathbf{I} & \frac{9}{ab}\mathbf{I} & 0 & 0 & \frac{9}{ab}\mathbf{I} & -\frac{9}{ab}\mathbf{I} & 0 & 0 \\
0 & 0 & \frac{3}{b}\mathbf{I} & 0 & 0 & 0 & -\frac{3}{b}\mathbf{I} & 0 \\
0 & 0 & \frac{9}{ab}\mathbf{I} & -\frac{9}{ab}\mathbf{I} & 0 & 0 & -\frac{9}{ab}\mathbf{I} & \frac{9}{ab}\mathbf{I}
\end{bmatrix} \qquad (16)$$

The correspondence of 16 ANCF absolute nodal coordinates and 16 Bezier surface control points in Eq.(12) is linear. An intuitive geometric description of this correspondence is shown in Figure 2. Figure 2 gives geometry of the $3\times 3$ Bezier surface expressed by 16 control points in Eq.(14) and the converted ANCF finite surface element obtained from Eq.(12) in terms of **e** expressed by these control points.

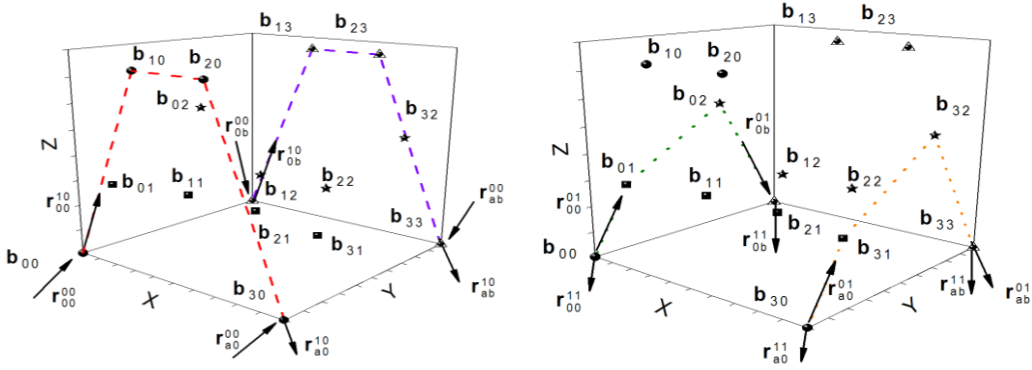

(a): The linear correspondence of absolute nodal coordinates and control points

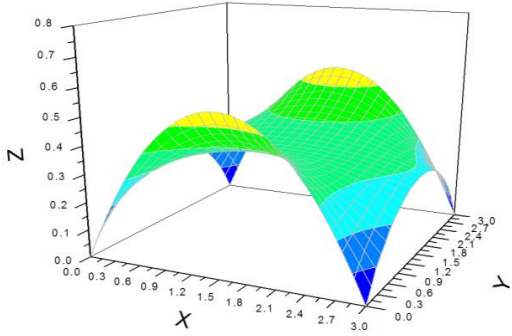

(b) The $3\times 3$ Bezier surface

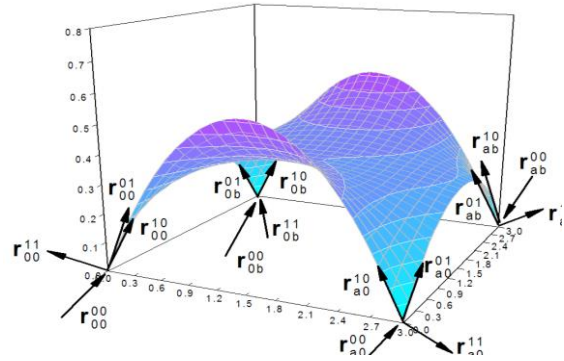

(c) The converted ANCF surface element

Figure 2 (a) is the geometric description of the linear correspondence of 16 ANCF absolute nodal coordinates and 16 Bezier surface control points. (b) is the $3\times 3$ Bezier surface in terms of 16 Bezier surface control points. (c) is the converted ANCF finite surface element in terms of 16 nodal coordinates expressed by 16 Bezier surface control points.

Eq.(12) utilizes Bezier surface control points to represent ANCF absolute nodal coordinates. It realizes a directed automated conversion from a known Bezier surface to an ANCF finite surface element. In addition, the directed automated inverse conversion from a known ANCF finite surface element to Bezier surface is necessary to further realize the bi-directed conversion. In Eq.(12), the blocks $\mathbf{A}$ and $\mathbf{B}$ in the diagonal block-partitioned matrix $\mathbf{T}$ are both invertible so $\mathbf{T}$ is invertible. The equation of inverse conversion can be derived from Eq.(12) as

$$\mathbf{b}_{ij} = \mathbf{T}^{-1}\mathbf{e} = \begin{bmatrix} \mathbf{A}^{-1} & \mathbf{0} \\ \mathbf{0} & \mathbf{B}^{-1} \end{bmatrix}\mathbf{e} \qquad (17)$$

where, $\mathbf{T}^{-1}$, $\mathbf{A}^{-1}$ and $\mathbf{B}^{-1}$ are the inverse matrices of $\mathbf{T}$, $\mathbf{A}$ and $\mathbf{B}$ in Eq.(12). The elements in $\mathbf{A}^{-1}$ and $\mathbf{B}^{-1}$ are given in Eq.(18) and Eq.(19).

Eq.(17) represents Bezier surface control points by ANCF absolute nodal coordinates. Given a known ANCF finite surface element, Eq.(17) and Bernstein polynomials are sufficient to convert the ANCF finite surface element to a Bezier surface. Therefore, the bi-directed automated conversion between Bezier surfaces and ANCF finite surface elements can be realized in terms of Eq.(12) and Eq.(17).

$$\mathbf{A}^{-1} = \begin{bmatrix} \mathbf{I} & 0 & 0 & 0 & 0 & 0 & 0 & 0 \\ \mathbf{I} & \frac{a}{3}\mathbf{I} & 0 & 0 & 0 & 0 & 0 & 0 \\ 0 & 0 & \mathbf{I} & 0 & 0 & 0 & 0 & 0 \\ 0 & 0 & \mathbf{I} & -\frac{a}{3}\mathbf{I} & 0 & 0 & 0 & 0 \\ \mathbf{I} & 0 & 0 & 0 & \frac{b}{3}\mathbf{I} & 0 & 0 & 0 \\ \mathbf{I} & \frac{a}{3}\mathbf{I} & 0 & 0 & \frac{b}{3}\mathbf{I} & \frac{ab}{9}\mathbf{I} & 0 & 0 \\ 0 & 0 & \mathbf{I} & 0 & 0 & 0 & \frac{b}{3}\mathbf{I} & 0 \\ 0 & 0 & \mathbf{I} & -\frac{a}{3}\mathbf{I} & 0 & 0 & \frac{b}{3}\mathbf{I} & -\frac{ab}{9}\mathbf{I} \end{bmatrix} \quad (18)$$

$$\mathbf{B}^{-1} = \begin{bmatrix} \mathbf{I} & 0 & 0 & 0 & 0 & 0 & 0 & 0 \\ \mathbf{I} & \frac{a}{3}\mathbf{I} & 0 & 0 & 0 & 0 & 0 & 0 \\ 0 & 0 & \mathbf{I} & 0 & 0 & 0 & 0 & 0 \\ 0 & 0 & \mathbf{I} & -\frac{a}{3}\mathbf{I} & 0 & 0 & 0 & 0 \\ \mathbf{I} & 0 & 0 & 0 & -\frac{b}{3}\mathbf{I} & 0 & 0 & 0 \\ \mathbf{I} & \frac{a}{3}\mathbf{I} & 0 & 0 & -\frac{b}{3}\mathbf{I} & -\frac{ab}{9}\mathbf{I} & 0 & 0 \\ 0 & 0 & \mathbf{I} & 0 & 0 & 0 & -\frac{b}{3}\mathbf{I} & 0 \\ 0 & 0 & \mathbf{I} & -\frac{a}{3}\mathbf{I} & 0 & 0 & -\frac{b}{3}\mathbf{I} & \frac{ab}{9}\mathbf{I} \end{bmatrix} \quad (19)$$

## 3. General conversion

The bi-directed automated conversion between ANCF finite surface element and $3\times 3$ Bezier surfaces is obtained in the above section. The general bi-directed automated conversion is existed between ANCF finite surface elements and $m\times n\,(m,n\leq 3,\,m,n\in N^{+})$ Bezier surfaces. This section represents this general conversion between ANCF and $m\times n\,(m,n<3,\,m,n\in N^{+})$ Bezier surfaces.

### 3.1 General transformation matrix

The principal rule of the conversion from $m\times n\,(m,n\leq 3,\,m,n\in N^{+})$ Bezier surfaces to ANCF finite surface elements is representing ANCF absolute nodal coordinates by Bezier surfaces control points. It can be deduced the conversion from Bezier surfaces control points to ANCF absolute nodal coordinates exists; the transformation matrix is linear in a unified form. Introduce $\mathbf{T}_g$ as the general transformation matrix. The equation of general conversion is

$$\mathbf{e} = \mathbf{T}_g \mathbf{b}_{ij} = \begin{bmatrix} \mathbf{A}_g & \mathbf{0} \\ \mathbf{0} & \mathbf{B}_g \end{bmatrix} \mathbf{b}_{ij} \qquad (20)$$

in which,

$$\mathbf{A}_g = \begin{bmatrix} \mathbf{I} & \mathbf{0} & \mathbf{0} & \mathbf{0} & \mathbf{0} & \mathbf{0} & \mathbf{0} & \mathbf{0} \\ -\frac{m}{a}\mathbf{I} & \frac{m}{a}\mathbf{I} & \mathbf{0} & \mathbf{0} & \mathbf{0} & \mathbf{0} & \mathbf{0} & \mathbf{0} \\ \mathbf{0} & \mathbf{0} & \mathbf{I} & \mathbf{0} & \mathbf{0} & \mathbf{0} & \mathbf{0} & \mathbf{0} \\ \mathbf{0} & \mathbf{0} & \frac{m}{a}\mathbf{I} & -\frac{m}{a}\mathbf{I} & \mathbf{0} & \mathbf{0} & \mathbf{0} & \mathbf{0} \\ -\frac{n}{b}\mathbf{I} & \mathbf{0} & \mathbf{0} & \mathbf{0} & \frac{n}{b}\mathbf{I} & \mathbf{0} & \mathbf{0} & \mathbf{0} \\ \frac{mn}{ab}\mathbf{I} & -\frac{mn}{ab}\mathbf{I} & \mathbf{0} & \mathbf{0} & -\frac{mn}{ab}\mathbf{I} & \frac{mn}{ab}\mathbf{I} & \mathbf{0} & \mathbf{0} \\ \mathbf{0} & \mathbf{0} & -\frac{n}{b}\mathbf{I} & \mathbf{0} & \mathbf{0} & \mathbf{0} & \frac{n}{3}\mathbf{I} & \mathbf{0} \\ \mathbf{0} & \mathbf{0} & -\frac{mn}{ab}\mathbf{I} & \frac{mn}{ab}\mathbf{I} & \mathbf{0} & \mathbf{0} & \frac{mn}{ab}\mathbf{I} & -\frac{mn}{ab}\mathbf{I} \end{bmatrix} \qquad (21)$$

$$\mathbf{B}_g = \begin{bmatrix} \mathbf{I} & \mathbf{0} & \mathbf{0} & \mathbf{0} & \mathbf{0} & \mathbf{0} & \mathbf{0} & \mathbf{0} \\ -\frac{m}{a}\mathbf{I} & \frac{m}{a}\mathbf{I} & \mathbf{0} & \mathbf{0} & \mathbf{0} & \mathbf{0} & \mathbf{0} & \mathbf{0} \\ \mathbf{0} & \mathbf{0} & \mathbf{I} & \mathbf{0} & \mathbf{0} & \mathbf{0} & \mathbf{0} & \mathbf{0} \\ \mathbf{0} & \mathbf{0} & \frac{m}{a}\mathbf{I} & -\frac{m}{a}\mathbf{I} & \mathbf{0} & \mathbf{0} & \mathbf{0} & \mathbf{0} \\ \frac{n}{b}\mathbf{I} & \mathbf{0} & \mathbf{0} & \mathbf{0} & -\frac{n}{b}\mathbf{I} & \mathbf{0} & \mathbf{0} & \mathbf{0} \\ -\frac{mn}{ab}\mathbf{I} & \frac{mn}{ab}\mathbf{I} & \mathbf{0} & \mathbf{0} & \frac{mn}{ab}\mathbf{I} & -\frac{mn}{ab}\mathbf{I} & \mathbf{0} & \mathbf{0} \\ \mathbf{0} & \mathbf{0} & \frac{n}{b}\mathbf{I} & \mathbf{0} & \mathbf{0} & \mathbf{0} & -\frac{n}{b}\mathbf{I} & \mathbf{0} \\ \mathbf{0} & \mathbf{0} & \frac{mn}{ab}\mathbf{I} & -\frac{mn}{ab}\mathbf{I} & \mathbf{0} & \mathbf{0} & -\frac{mn}{ab}\mathbf{I} & \frac{mn}{ab}\mathbf{I} \end{bmatrix} \qquad (22)$$

where, $m$ and $n$ are from the definition of $m \times n$ Bezier surfaces representing the degrees of Bezier function; $a$ and $b$ are from the shape functions of ANCF finite surface element standing for length and width of the ANCF finite surface element.

### 3.2 A conversion case

Eq.(20) and Eq.(6) realize the conversion from $m \times n$ ($m, n \leq 3$, $m, n \in N^+$) Bezier surfaces to ANCF finite surface elements. In this section, a case of the conversion from $3 \times 2$ Bezier surfaces to ANCF finite surface elements is elaborated. Other conversion cases can be analogously handled.

Bezier surfaces control points are utilized to represent ANCF absolute nodal coordinates in Eq.(20). In $3 \times 2$ Bezier surfaces, there are 12 control points, 4 of which lying on the surface represent 4 position vectors of ANCF finite surface element and the rest 8 of which represent 12 gradient vectors of ANCF finite surface element. 4 of the control points $\mathbf{b}_{01}, \mathbf{b}_{11}, \mathbf{b}_{31}, \mathbf{b}_{21}$ are utilized twice. The geometric explanation of this representing process refers to Figure 4.

Figure 3 shows a $3 \times 2$ Bezier surface over the domain $(u,v) \in [0,1] \times [0,1]$ and the corresponding 12 control points.

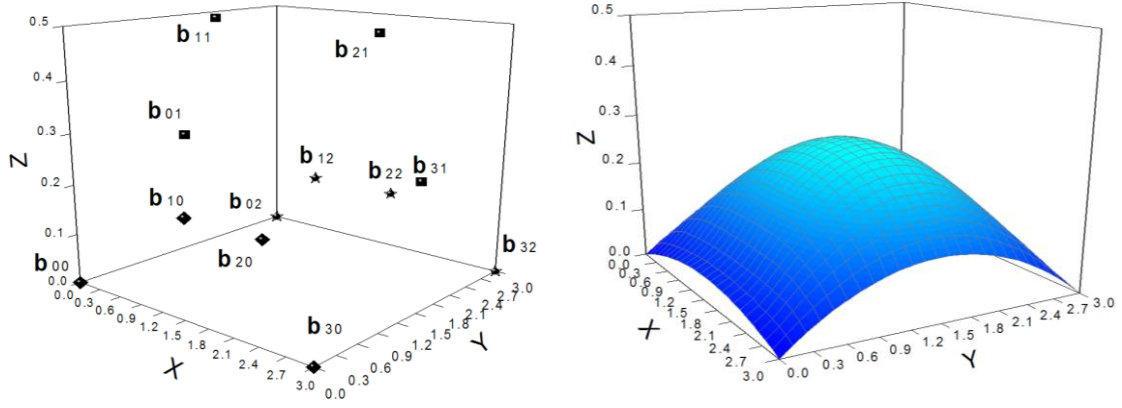

Figure 3 A $3 \times 2$ Bezier surface and its corresponding 12 control points

Converting this Bezier surface to the corresponding ANCF finite surface element depends on the Eq.(20) and Eq.(6). The vector of control points $\mathbf{b}_{ij}$ utilized in the Eq.(20) is Eq.(23). It is proved the expression of corresponding converted ANCF finite surface element is same as the expression of $3 \times 2$ Bezier surface. Other conversions from $m \times n$ ($m, n \leq 3$, $m, n \in N^+$) Bezier surfaces to ANCF finite surface elements come to the same conclusion. The general conversion equation works in the conversion from $m \times n$ ($m, n \leq 3$, $m, n \in N^+$) Bezier surfaces to ANCF finite surface elements. In addition to algebraic representation of this conversion process, the geometric explanation of the relationship between Bezier surfaces control points and ANCF absolute nodal coordinates and the comparison of geometry before and after conversion is given in the Figure 4. One can get the analogous geometric explanation on the other conversion cases.

$$\mathbf{b}_{ij} = [\mathbf{b}_{00}, \mathbf{b}_{10}, \mathbf{b}_{30}, \mathbf{b}_{20}, \mathbf{b}_{01}, \mathbf{b}_{11}, \mathbf{b}_{31}, \mathbf{b}_{21}, \mathbf{b}_{02}, \mathbf{b}_{12}, \mathbf{b}_{32}, \mathbf{b}_{22}, \mathbf{b}_{01}, \mathbf{b}_{11}, \mathbf{b}_{31}, \mathbf{b}_{21}]^T \qquad (23)$$

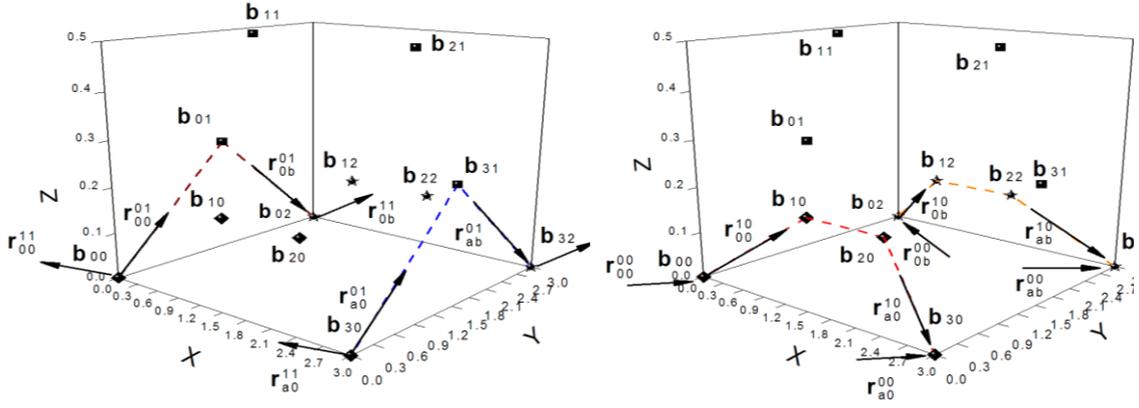

(a): Relationship between Bezier surfaces control points and ANCF absolute nodal coordinates

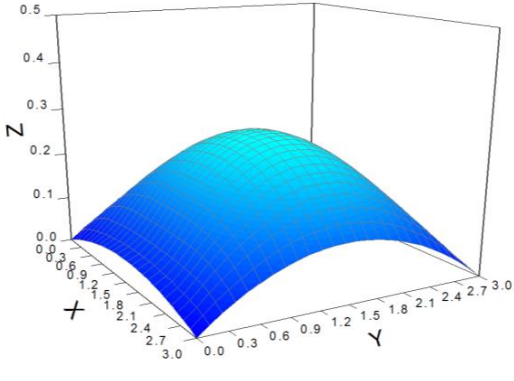

(b) The $3\times 2$ Bezier surface

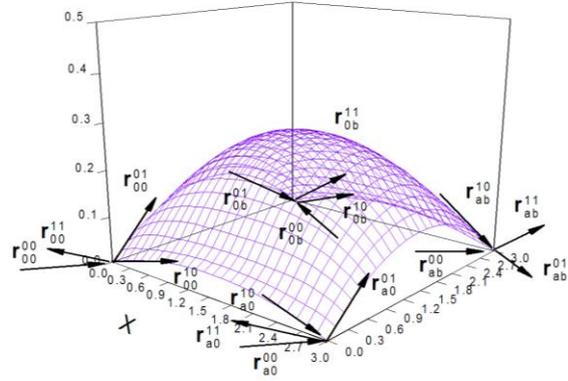

(c) The converted ANCF surface element

Figure 4 (a) is the geometric explanation of relationship between 12 Bezier surfaces control points and 16 ANCF absolute nodal coordinates. (b) is the $3\times 2$ Bezier surface in terms of 12 Bezier surface control points. (c) is the converted ANCF finite surface element in terms of 16 nodal coordinates expressed by 12 Bezier surface control points.

The functionality of the transformation matrix in Eq.(20) leading the conversion from $m \times n\ (m,n \leq 3,\ m,n \in N^{+})$ Bezier surfaces to ANCF finite surface elements is proved both algebraically and geometrically. Next we deduce equations of the inverse conversion from a known ANCF finite surface element to Bezier surface, a necessity to realize the bi-directed automated conversion between them. It can be deduced from Eq.(20). In Eq.(20), the blocks $\mathbf{A}_g$ and $\mathbf{B}_g$ in the general transformation matrix $\mathbf{T}_g$ are both invertible therefore $\mathbf{T}_g$ is invertible. The inverse conversion equation has the form as follows

$$\mathbf{b}_{ij} = \mathbf{T}_g^{-1}\mathbf{e} = \begin{bmatrix} \mathbf{A}_g^{-1} & \mathbf{0} \\ \mathbf{0} & \mathbf{B}_g^{-1} \end{bmatrix}\mathbf{e} \tag{24}$$

where, $\mathbf{T}_g^{-1}$, $\mathbf{A}_g^{-1}$ and $\mathbf{B}_g^{-1}$ is the inverse matrices of $\mathbf{T}_g$, $\mathbf{A}_g$ and $\mathbf{B}_g$ in Eq.(20). The elements in $\mathbf{A}_g^{-1}$ and $\mathbf{B}_g^{-1}$ are shown in Eq.(25) and Eq.(26).

$$\mathbf{A}_g^{-1} = \begin{bmatrix} \mathbf{I} & 0 & 0 & 0 & 0 & 0 & 0 & 0 \\ \mathbf{I} & \frac{a}{m}\mathbf{I} & 0 & 0 & 0 & 0 & 0 & 0 \\ 0 & 0 & \mathbf{I} & 0 & 0 & 0 & 0 & 0 \\ 0 & 0 & \mathbf{I} & -\frac{a}{m}\mathbf{I} & 0 & 0 & 0 & 0 \\ \mathbf{I} & 0 & 0 & 0 & \frac{b}{n}\mathbf{I} & 0 & 0 & 0 \\ \mathbf{I} & \frac{a}{m}\mathbf{I} & 0 & 0 & \frac{b}{n}\mathbf{I} & \frac{ab}{mn}\mathbf{I} & 0 & 0 \\ 0 & 0 & \mathbf{I} & 0 & 0 & 0 & \frac{b}{n}\mathbf{I} & 0 \\ 0 & 0 & \mathbf{I} & -\frac{a}{m}\mathbf{I} & 0 & 0 & \frac{b}{n}\mathbf{I} & -\frac{ab}{mn}\mathbf{I} \end{bmatrix} \quad (25)$$

$$\mathbf{B}_g^{-1} = \begin{bmatrix} \mathbf{I} & 0 & 0 & 0 & 0 & 0 & 0 & 0 \\ \mathbf{I} & \frac{a}{m}\mathbf{I} & 0 & 0 & 0 & 0 & 0 & 0 \\ 0 & 0 & \mathbf{I} & 0 & 0 & 0 & 0 & 0 \\ 0 & 0 & \mathbf{I} & -\frac{a}{m}\mathbf{I} & 0 & 0 & 0 & 0 \\ \mathbf{I} & 0 & 0 & 0 & -\frac{b}{n}\mathbf{I} & 0 & 0 & 0 \\ \mathbf{I} & \frac{a}{m}\mathbf{I} & 0 & 0 & -\frac{b}{n}\mathbf{I} & -\frac{ab}{mn}\mathbf{I} & 0 & 0 \\ 0 & 0 & \mathbf{I} & 0 & 0 & 0 & -\frac{b}{n}\mathbf{I} & 0 \\ 0 & 0 & \mathbf{I} & -\frac{a}{m}\mathbf{I} & 0 & 0 & -\frac{b}{n}\mathbf{I} & \frac{ab}{mn}\mathbf{I} \end{bmatrix} \quad (26)$$

The conversion from an ANCF finite surface element to a Bezier surface can be obtained in terms of Eq.(24) and Bernstein polynomials. Therefore, the general bi-directed automated conversion between Bezier surfaces and ANCF finite surface elements can be realized in terms of Eq.(20) and Eq.(24).

### 3.3 Independence of position and gradient variables

The proposed equations Eq.(20) and Eq.(24) have a determinative functionality on conversion between Bezier surfaces and ANCF finite surface elements. It promotes the integration between computer aided design and analysis. 3-order ANCF finite surface elements utilized in this literature are better in the dynamic analysis compared to lower-order ones because it is stronger to describe the dynamic behaviors of the multibody system. In comparison, the Bezier surface represented by Bezier functions of degree $m \times n$ in CAD systems can also be represented by Bezier functions of degree $(m+\Delta m) \times (n+\Delta n)$, $\Delta m, \Delta n \in N$. In this process the order and control points of Bezier increase without change of geometry. In this process, lower-degrees $m \times n\,(m,n<3, m,n \in N^+)$ Bezier functions describe the same geometry as higher degree ones. We may call the lowest-degrees to describe the same geometry as the optimal degrees. Thus, conditional converted Bezier surfaces from ANCF surfaces can lower their degrees by the inverse process. This inverse process enhances the ability of data storage in the conversion process because the optimal degrees Bezier surfaces have less control points and simpler function forms.

The control points of a $m \times n\,(m,n<3, m,n \in N^+)$ Bezier surface is less than 16 so the 16 converted ANCF absolute nodal coordinates are not independent. Suppose the degree of the Bezier surface is $3 \times 2$, the 16 dependent converted ANCF absolute nodal coordinates have the relationship as follows

$$\begin{cases} \mathbf{r}_{0b}^{01} = \dfrac{2}{b}\mathbf{r}_{0b}^{00} - \dfrac{2}{b}\mathbf{r}_{00}^{00} - \mathbf{r}_{00}^{01} \\[6pt] \mathbf{r}_{ab}^{01} = \dfrac{2}{b}\mathbf{r}_{ab}^{00} - \dfrac{2}{b}\mathbf{r}_{a0}^{00} - \mathbf{r}_{a0}^{01} \\[6pt] \mathbf{r}_{0b}^{11} = \dfrac{2}{b}\mathbf{r}_{0b}^{10} - \dfrac{2}{b}\mathbf{r}_{00}^{10} - \mathbf{r}_{00}^{11} \\[6pt] \mathbf{r}_{ab}^{11} = \dfrac{2}{b}\mathbf{r}_{ab}^{10} - \dfrac{2}{b}\mathbf{r}_{00}^{01} - \mathbf{r}_{a0}^{11} \end{cases} \tag{27}$$

In Eq.(27), 4 of the converted ANCF absolute nodal coordinates are not independent and can be represented by the other 12 nodal coordinates. If the known ANCF finite surface elements being converted satisfy the dependent relationship in Eq.(27), the optimal-degrees of Bezier surface obtained should be a $3 \times 2$ one, but the result is actually a $3 \times 3$ Bezier surface from Eq.(24) and Eq.(1) provided that computers are not told about the dependent relationship. But if given the analogous dependent relationships which can be obtained from the conversion from Bezier surfaces to ANCF surface elements, the inverse conversion will directly get optimal-degrees Bezier surfaces satisfying the same geometry. This dependence of ANCF surface elements variables improves the ability to store data of converted Bezier surfaces. The $m \times n\,(m,n<3, m,n \in N^+)$ Bezier surfaces can also be converted to lower-order ANCF surface elements, but it constrains the dynamics analysis ability of them so this process is not illustrated.

## 4. Conversion between B-spline surface and ANCF finite surface elements

The general conversions between ANCF finite surface elements and Bezier surfaces are established. B-spline surface representation is another common methodology to describe surface in CAD systems. B-spline surfaces are a generalization of Bezier surfaces and have more desired properties on geometric description i.e., local modification scheme. Therefore, successes on the conversion between B-spline surface and ANCF finite surface elements can further boost the integration between CAD and CAA.

### 4.1 B-spline surface representation

This section gives a brief introduction on B-spline surface basis functions and B-spline geometry. It introduces several fundamental notations utilized to derive the conversion equation between ANCF finite surface elements and B-spline surfaces in following sections. For details on B-spline surfaces, refer to *The NURBS Book* [19].

A B-spline surface of degrees $k \times l$ is defined as follows

$$p(u,v) = \sum_{i=0}^{m}\sum_{j=0}^{n} d_{ij} B_{i,k}(u) B_{j,l}(v) \tag{28}$$

where $d_{ij}(i=0,\cdots,m; j=0,\cdots,n)$ represents the coordinate positions of a set of $(m+1)\times(n+1)$ control points. $B_{i,k}(u)$ and $B_{j,l}(v)$ are the degrees $k$ and $l$ B-spline basis functions defined on the knot vectors $\mathbf{U}$ and $\mathbf{V}$ as follows

$$\mathbf{U} = \{u_0, u_1, \cdots, u_{m+k+1}\} \tag{29}$$

$$\mathbf{V} = \{v_0, v_1, \cdots, v_{n+l+1}\} \tag{30}$$

The $i$-th $(i=0,\cdots,m)$ and $j$-th $(j=0,\cdots,n)$ B-spline basis functions in $B_{i,k}(u)$ and $B_{j,l}(v)$ can be defined recursively as

$$\begin{cases} B_{i,0}(u) = \begin{cases} 1 & \text{if } u_i \leq u \leq u_{i+1} \\ 0 & \text{otherwise} \end{cases} \\ B_{i,k}(u) = \dfrac{u-u_i}{u_{i+k}-u_i} B_{i,j-1}(u) + \dfrac{u_{i+k+1}-u}{u_{i+k+1}-u_{i+1}} B_{i+1,k-1}(u) \\ \text{define } \dfrac{0}{0} = 0 \end{cases} \tag{31}$$

$$\begin{cases} B_{j,0}(v) = \begin{cases} 1 & \text{if } v_j \leq v \leq v_{j+1} \\ 0 & \text{otherwise} \end{cases} \\ B_{j,l}(v) = \dfrac{v-v_j}{v_{j+l}-v_j} B_{j,l-1}(v) + \dfrac{v_{j+l+1}-v}{v_{j+l+1}-v_{j+1}} B_{j+1,l-1}(v) \\ \text{define } \dfrac{0}{0} = 0 \end{cases} \tag{32}$$

Figure 5 shows a $3\times 3$ B-spline surface and its 24 corresponding control points $d_{ij}(i=0,...,3; j=0,...,5)$.

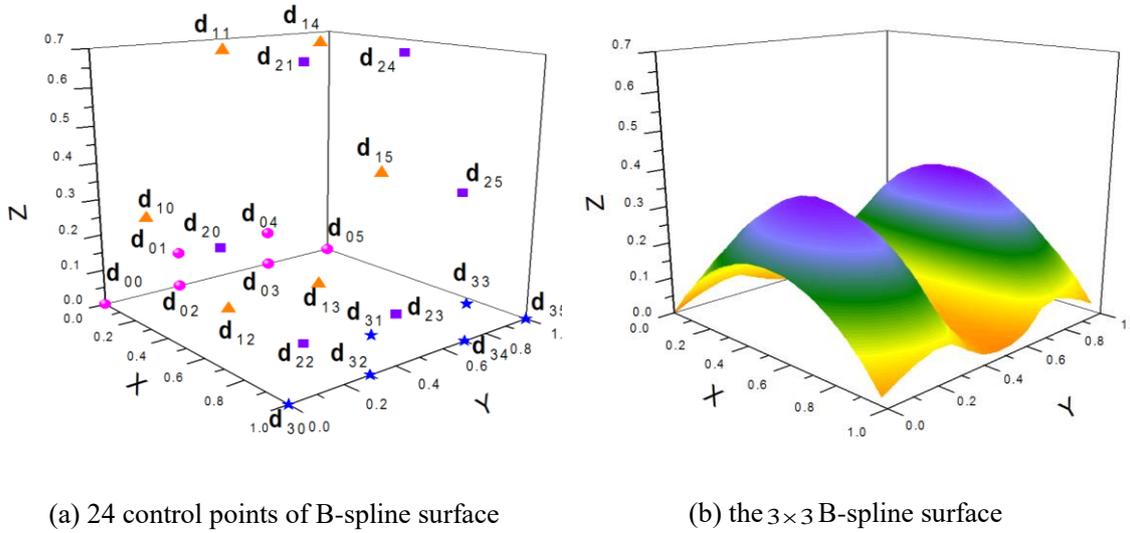

(a) 24 control points of B-spline surface  (b) the $3\times3$ B-spline surface

Figure 5: $3\times3$ B-spline surface with degrees $3\times3$ for knot vector $\mathbf{U}=\{0,0,0,0,1,1,1,1\}$ and $\mathbf{V}=\{0,0,0,0,0.4,0.7,1,1,1,1\}$

The surface segment defined on the non-zero area $u_e \leq u \leq u_{e+1}$, $v_f \leq v \leq v_{f+1}$ is only related to partial control points in De Boor's algorithm, so Eq.(28) is simplified as

$$p(u,v) = \sum_{i=e-k}^{e} \sum_{j=f-l}^{f} d_{ij} B_{i,k}(u) B_{j,l}(v) \tag{33}$$

where,

$$u \in [u_e, u_{e+1}] \subset [u_k, u_{m+1}], v \in [v_f, v_{f+1}] \subset [v_l, v_{n+1}] \tag{34}$$

The conversion between B-spline surfaces and ANCF finite surfaces elements can be realized because B-spline surfaces can be obtained by Bezier surfaces inter-connection. This connection process is realized by inserting knots repeatedly using the blossoming formula[21] to increase the knot multiplicity. If the knot multiplicity of partial control points are increased to $m+1$ and $n+1$ ($m,n$ are the degrees of the B-spline surface), and the multiplicity at other control points is $m$ and $n$, the B-spline surface is a composite Bezier surfaces. Bezier surfaces can be converted to ANCF finite surface elements, so a B-spline surface can be converted to an ANCF finite surface element and the converted ANCF finite surface element is a montage of a set of ANCF finite surface elements converted from Bezier surfaces. This semantics of converting B-spline surfaces to ANCF finite surface element is very intuitive. But a more efficient conversion is found between ANCF finite surface elements and $k\times l$ ($k,l \leq 3$, $k,l \in N^+$) B-spline surfaces. The conversion between ANCF finite surface elements and $3\times3$ B-spline surfaces is first derived and the conversion between ANCF and $k\times l$ ($k,l < 3$, $k,l \in N^+$) B-spline surfaces are proposed afterwards.

Suppose a $3\times3$ B-spline surface defined in non-zero area $u_e \leq u \leq u_{e+1}$, $v_f \leq v \leq v_{f+1}$. The function of this surface is given from Eq.(33) as follows

$$p(u,v) = \sum_{i=e-3}^{e} \sum_{j=f-3}^{f} d_{ij} B_{i,3}(u) B_{j,3}(v) \tag{35}$$

where $B_{i,3}(u)$ and $B_{j,3}(v)$ are defined on the knot vectors $\mathbf{U}_e$ and $\mathbf{V}_f$ as

$$\mathbf{U}_e = \{u_{e-2}, u_{e-1}, u_e, u_{e+1}, u_{e+2}, u_{e+3}\} \tag{36}$$

$$\mathbf{V}_f = \{v_{f-2}, v_{f-1}, v_f, v_{f+1}, v_{f+2}, v_{f+3}\} \tag{37}$$

From Eq.(31) and Eq.(32), the B-spline basis functions can be expanded into a non-recursive form

$$\begin{cases} B_{\alpha-3,3}(\lambda) = \dfrac{F_1^3}{H_{1,-2}H_{1,-1}H_{1,0}} \\[6pt] B_{\alpha-2,3}(\lambda) = \dfrac{F_1^2 G_{-2}}{H_{1,-2}H_{1,-1}H_{1,0}} + \dfrac{F_1 F_2 G_{-1}}{H_{2,-1}H_{1,-1}H_{1,0}} + \dfrac{F_2^2 G_0}{H_{2,-1}H_{2,0}H_{1,0}} \\[6pt] B_{\alpha-1,3}(\lambda) = \dfrac{F_1 G_{-1}^2}{H_{2,-1}H_{1,-1}H_{1,0}} + \dfrac{F_2 G_0 G_{-1}}{H_{2,-1}H_{2,0}H_{1,0}} + \dfrac{F_3 G_0^2}{H_{3,0}H_{2,0}H_{1,0}} \\[6pt] B_{\alpha,3}(\lambda) = \dfrac{G_0^3}{H_{3,0}H_{2,0}H_{1,0}} \end{cases} \tag{38}$$

where $\lambda = u, v$, $F_\beta$, $G_m$ and $H_{\beta,\gamma}$ are defined as

$$\begin{cases} F_\beta = \lambda_{\alpha+\beta} - \lambda \\ G_m = \lambda - \lambda_{\alpha+\beta} \\ H_{\beta,\gamma} = \lambda_{\alpha+\beta} - \lambda_{\alpha+\gamma} \end{cases} \tag{39}$$

in which,

$$\begin{cases} \alpha = e, \beta = m, \gamma = n & \text{if } \lambda = u \\ \alpha = f, \beta = p, \gamma = q & \text{if } \lambda = v \end{cases} \tag{40}$$

It can be shown from Eq. (38) ~ Eq. (40) that $K_{\alpha-3,3}(\lambda_{i+1}) = 0$ and $K_{\alpha,3}(\lambda_i) = 0$. It implies 4 points $\mathbf{p}(u_i, v_j), \mathbf{p}(u_i, v_{j+1}), \mathbf{p}(u_{i+1}, v_j)$ and $\mathbf{p}(u_{i+1}, v_{j+1})$ of the surface are not controlled by control points $\mathbf{d}_{i,j}, \mathbf{d}_{i,j-3}, \mathbf{d}_{i-3,j}$ and $\mathbf{d}_{i-3,j-3}$, respectively, which allows for further simplification on equations utilizing these surface end-points.

**4.2 Conversion between ANCF and B-spline surfaces**

The conversion of the B-spline surface to an ANCF finite surface element requires finding the corresponding relationship between control points and ANCF nodal coordinates. This corresponding relationship is obtained by Eq.(41)~Eq.(43). The nodal coordinates of corresponding ANCF element converted from a B-spline surface can be obtained Eq.(35):

$$\begin{cases} \mathbf{r}_{uv}^{00} = \mathbf{p}(u_a, v_b) \\ \mathbf{r}_{uv}^{10} = \dfrac{\partial \mathbf{p}(u,v)}{\partial u} \dfrac{\partial u}{\partial \xi} \dfrac{\partial \xi}{\partial x}\Big|_{u=u_a, v=v_b} \\ \mathbf{r}_{uv}^{01} = \dfrac{\partial \mathbf{p}(u,v)}{\partial v} \dfrac{\partial v}{\partial \eta} \dfrac{\partial \eta}{\partial y}\Big|_{u=u_a, v=v_b} \\ \mathbf{r}_{uv}^{11} = \dfrac{\partial \mathbf{p}(u,v)}{\partial u \partial v} \dfrac{\partial u}{\partial \xi} \dfrac{\partial \xi}{\partial x} \dfrac{\partial v}{\partial \eta} \dfrac{\partial \eta}{\partial y}\Big|_{u=u_a, v=v_b} \end{cases} \quad (41)$$

where, $a = i, i+1; b = j, j+1$. $x$ is the function of $\xi$; $\xi$ is the function of $u$. $y$ is the function of $\eta$; $\eta$ is the function of $v$. $\mathbf{r}_{uv}^{ij}$ ($i = 0,1; j = 0,1$) are functions of $u$ and $v$. The functions $x(\xi), y(\eta), \xi(u)$ and $\eta(v)$ are given in Eq.(42) and Eq.(43).

$$\begin{cases} x = \xi a \\ y = \eta b \end{cases} \quad (42)$$

Eq.(42) shows the relationship of the parameters $x, y$ in the ANCF finite surface element and the length $a$ and width $b$ of the B-spline surface.

$$\begin{cases} \xi = (u - u_i)/(u_{i+1} - u_i) \\ \eta = (v - v_i)/(v_{i+1} - v_i) \end{cases} \quad (43)$$

Eq.(43) shows the relationship of variables $\xi, \eta$ of the ANCF finite surface element and variables $u, v$ of B-spline parametric surface.

From Eq.(41) ~ Eq.(43), the 16 nodal coordinates of the ANCF finite surface element represented by the B-spline control points are obtained. The equations refer to Appendix A. From these equations, the transformation matrix from a $3 \times 3$ B-spline surface to an ANCF finite surface element is derived. Eq.(44) gives the transformation matrix $\mathbf{\Psi}$ and the conversion equation.

$$\mathbf{e} = \mathbf{\Psi} \mathbf{d}_{ij} = \begin{bmatrix} \mathbf{\Psi}_{11} & \mathbf{\Psi}_{12} & \mathbf{\Psi}_{13} & \mathbf{\Psi}_{14} \\ \mathbf{\Psi}_{21} & \mathbf{\Psi}_{22} & \mathbf{\Psi}_{23} & \mathbf{\Psi}_{24} \\ \mathbf{\Psi}_{31} & \mathbf{\Psi}_{32} & \mathbf{\Psi}_{33} & \mathbf{\Psi}_{34} \\ \mathbf{\Psi}_{41} & \mathbf{\Psi}_{42} & \mathbf{\Psi}_{43} & \mathbf{\Psi}_{44} \end{bmatrix} \mathbf{d}_{ij} \quad (44)$$

in which, the blocks $\mathbf{\psi}_{ij}$ ($i, j = 0, 1, 2, 3$) refer to appendix B. The vector of B-spline control points is defined as

$$\mathbf{d}_{ij} = \big[ \mathbf{d}_{e-3,f-3}, \mathbf{d}_{e-3,f-2}, \mathbf{d}_{e-3,f-1}, \mathbf{d}_{e-3,f}, \mathbf{d}_{e-2,f-3}, \mathbf{d}_{e-2,f-2}, \mathbf{d}_{e-2,f-1}, \mathbf{d}_{e-2,f}, \\ \mathbf{d}_{e-1,f-3}, \mathbf{d}_{e-1,f-2}, \mathbf{d}_{e-1,f-1}, \mathbf{d}_{e-1,f}, \mathbf{d}_{e,f-3}, \mathbf{d}_{e,f-2}, \mathbf{d}_{e,f-1}, \mathbf{d}_{e,f} \big] \quad (45)$$

A $3 \times 3$ B-spline surface can be converted to an ANCF finite surface element in terms of Eq.(44) and Eq.(6). So the displacement field described by the 16 independent ANCF nodal coordinates or the 16 independent B-spline surface control points is the same. Consider them as the two bases of the displacement field with the basis transformation matrix $\mathbf{\Psi}$ hence the inverse conversion from ANCF displacement field to B-spline geometric field exists. The inverse conversion is given by Eq.(46).

$$\mathbf{d}_{ij} = \mathbf{\Psi}^{-1} \mathbf{e} \quad (46)$$

The conversion equation between a B-spline surface of degree $k \times l$ ($k, l < 3, k, l \in N^+$) and an ANCF finite surface element can be obtained in terms of the given non-recursive B-spline basis functions and Eq.(41)~Eq.(43). For instance, a $3 \times 2$ B-spline surface defined on the knot vectors

$$U_e = \{u_{e-2}, u_{e-1}, u_e, u_{e+1}, u_{e+2}, u_{e+3}\} \tag{47}$$

$$V_f = \{v_{f-1}, v_f, v_{f+1}, v_{f+2}\} \tag{48}$$

the function of which is

$$p(u,v) = \sum_{i=e-3}^{e} \sum_{j=f-2}^{f} d_{ij} B_{i,3}(u) B_{j,2}(v) \tag{49}$$

where $B_{i,3}(u)$ can be obtained by Eq.(38) and the $B_{j,2}(v)$ can be obtained by Eq.(50).

$$\begin{cases} B_{\alpha-2,2}(\lambda) = \dfrac{F_1^2}{H_{1,-1}} \\ B_{\alpha-1,2}(\lambda) = \dfrac{F_1 G_{-1}}{H_{1,0} H_{1,-1}} + \dfrac{F_2 G_0}{H_{2,0} H_{1,0}} \\ B_{\alpha,2}(\lambda) = \dfrac{G_0^2}{H_{2,0}} \end{cases} \tag{50}$$

where, $F, G$ and $H$ refers to Eq.(39). The values of B-spline basis functions and corresponding derivatives used in the conversion equation are non-recursively represented as

$$\begin{cases} B_{\alpha-2,2}(\lambda_i) = \dfrac{H_{1,0}^2}{H_{1,-1}} \\ B_{\alpha-1,2}(\lambda_i) = \dfrac{H_{0,-1}}{H_{1,-1}} \\ B'_{\alpha-2,2}(\lambda_i) = -\dfrac{2 H_{1,0}}{H_{1,-1}} \\ B'_{\alpha-1,2}(\lambda_i) = \dfrac{2}{H_{1,-1}} \\ B_{\alpha-1,2}(\lambda_{i+1}) = \dfrac{H_{2,1}}{H_{2,0}} \\ B_{\alpha,2}(\lambda_{i+1}) = \dfrac{H_{1,0}^2}{H_{2,0}} \\ B'_{\alpha-1,2}(\lambda_{i+1}) = -\dfrac{2}{H_{2,0}} \\ B'_{\alpha,2}(\lambda_{i+1}) = \dfrac{2 H_{1,0}}{H_{2,0}} \end{cases} \tag{51}$$

The conversion equation between $3 \times 2$ B-spline surfaces to ANCF finite surface elements can be derived from Eq.(49)~Eq.(51) in an analogous way as we did in conversion on $3 \times 3$ B-spline surfaces. At last, $K_{\alpha-1,1}(\lambda) = F_1$ and $K_{\alpha,1}(\lambda) = G_0$ are additional required conditions in conversion from $k \times l (k, l \leq 3, k\ or\ l = 1)$ B-spline surfaces to ANCF finite surface elements. The conversion between $k \times l\ (k, l \leq 3, k, l \in N^+)$ B-spline surfaces and ANCF finite surface elements based on non-recursive basis functions are proposed. This approach has a higher efficiency compared to converting firstly B-spline surfaces to composite Bezier surfaces by inserting knot and then converting to corresponding ANCF finite surface elements afterwards. The bi-directed conversion between $k \times l\ (k, l \leq 3, k, l \in N^+)$ B-spline surfaces and ANCF finite surface elements can be realized automatically by computers in one step.

## 5. Conclusions

This paper presents general conversion equations between ANCF finite surface elements and $m \times n\ (m, n \leq 3, m, n \in N^+)$ Bezier, $k \times l\ (k, l \leq 3, k, l \in N^+)$ B-spline surfaces. It extends our previous work on ANCF cable elements and B-spline curves to 2-dimensional thin plate elements and B-spline surfaces. The conversion equations established in this study are successfully applied to conversions of different degrees ($\leq 3 \times 3$) of Bezier and B-spline surfaces.

The geometric invariance of the ANCF displacement field before and after the conversion between ANCF finite surface elements and different degrees of Bezier, B-spline surfaces help establish the correspondence of ANCF absolute nodal coordinates and control points of Bezier, B-spline surfaces. In addition, the existence of general forms of linear transformation matrices unifies the conversion equations between ANCF and Bezier, B-spline surfaces. These conversions can be realized in one-step in terms of these linear transformation matrices so the conversion efficiency is acceptable. The independence of ANCF nodal coordinates is considered as a factor of affecting the conversion efficiency. Conditional ANCF finite surface elements satisfying the dependency of nodal coordinates can be converted to optimal-degrees Bezier and B-spline surfaces expressed by less control points and simpler function form. This boosts the efficiency and ability to control and store data in computers during the conversion process.

**Acknowledgements**

This research is supported by National Natural Science Foundation of China (Grant No. 11172076 ) and by Science and Technology Innovation Talent Foundation of Harbin (2012RFLXG020).

# Appendix A

$$
\begin{cases}
\mathbf{r}_{00}^{00} = \mathbf{p}(u_i, v_j) = \sum_{i=e-3}^{e-1} \sum_{j=f-3}^{f-1} \mathbf{d}_{i,j} B_{i,3}(u_i) B_{j,3}(v_j) \\[4pt]
\mathbf{r}_{00}^{10} = \dfrac{\partial \mathbf{p}(u,v)}{\partial u}\bigg|_{u=u_i, v=v_j} = \dfrac{1}{a}(u_{i+1}-u_i) \sum_{i=e-3}^{e-1} \sum_{j=f-3}^{f-1} \mathbf{d}_{i,j} (B_{i,3})_{,u}\big|_{u=u_i} B_{j,3}(v_j) \\[4pt]
\mathbf{r}_{00}^{01} = \dfrac{\partial \mathbf{p}(u,v)}{\partial v}\bigg|_{u=u_i, v=v_j} = \dfrac{1}{b}(v_{j+1}-v_j) \sum_{i=e-3}^{e-1} \sum_{j=f-3}^{f-1} \mathbf{d}_{i,j} B_{i,3}(u_i) (B_{j,3})_{,v}\big|_{v=v_j} \\[4pt]
\mathbf{r}_{00}^{11} = \dfrac{\partial \mathbf{p}(u,v)}{\partial u \partial v}\bigg|_{u=u_i, v=v_j} = \dfrac{1}{a}\dfrac{1}{b}(u_{i+1}-u_i)(v_{j+1}-v_j) \sum_{i=e-3}^{e-1} \sum_{j=f-3}^{f-1} \mathbf{d}_{i,j} (B_{i,3})_{,u}\big|_{u=u_i} (B_{j,3})_{,v}\big|_{v=v_j} \\[4pt]
\mathbf{r}_{a0}^{00} = \mathbf{p}(u_{i+1}, v_j) = \sum_{i=e-2}^{e} \sum_{j=f-3}^{f-1} \mathbf{d}_{i,j} B_{i,3}(u_{i+1}) B_{j,3}(v_j) \\[4pt]
\mathbf{r}_{a0}^{10} = \dfrac{\partial \mathbf{p}(u,v)}{\partial u}\bigg|_{u=u_{i+1}, v=v_j} = \dfrac{1}{a}(u_{i+1}-u_i) \sum_{i=e-2}^{e} \sum_{j=f-3}^{f-1} \mathbf{d}_{i,j} (B_{i,3})_{,u}\big|_{u=u_{i+1}} B_{j,3}(v_j) \\[4pt]
\mathbf{r}_{a0}^{01} = \dfrac{\partial \mathbf{p}(u,v)}{\partial v}\bigg|_{u=u_{i+1}, v=v_j} = \dfrac{1}{b}(v_{j+1}-v_j) \sum_{i=e-3}^{e} \sum_{j=f-3}^{f-1} \mathbf{d}_{i,j} B_{i,3}(u_{i+1}) (B_{j,3})_{,v}\big|_{v=v_j} \\[4pt]
\mathbf{r}_{a0}^{11} = \dfrac{\partial \mathbf{p}(u,v)}{\partial u \partial v}\bigg|_{u=u_{i+1}, v=v_j} = \dfrac{1}{a}\dfrac{1}{b}(u_{i+1}-u_i)(v_{j+1}-v_j) \sum_{i=e-3}^{e} \sum_{j=f-3}^{f-1} \mathbf{d}_{i,j} (B_{i,3})_{,u}\big|_{u=u_{i+1}} (B_{j,3})_{,v}\big|_{v=v_j} \\[4pt]
\mathbf{r}_{0b}^{00} = \mathbf{p}(u_i, v_{j+1}) = \sum_{i=e-3}^{e-1} \sum_{j=f-2}^{f} \mathbf{d}_{i,j} B_{i,3}(u_i) B_{j,3}(v_{j+1}) \\[4pt]
\mathbf{r}_{0b}^{10} = \dfrac{\partial \mathbf{p}(u,v)}{\partial u}\bigg|_{u=u_i, v=v_{j+1}} = \dfrac{1}{a}(u_{i+1}-u_i) \sum_{i=e-3}^{e-1} \sum_{j=f-2}^{f} \mathbf{d}_{i,j} (B_{i,3})_{,u}\big|_{u=u_i} B_{j,3}(v_{j+1}) \\[4pt]
\mathbf{r}_{0b}^{01} = \dfrac{\partial \mathbf{p}(u,v)}{\partial v}\bigg|_{u=u_i, v=v_{j+1}} = \dfrac{1}{b}(v_{j+1}-v_j) \sum_{i=e-3}^{e-1} \sum_{j=f-2}^{f} \mathbf{d}_{i,j} B_{i,3}(u_i) (B_{j,3})_{,v}\big|_{v=v_{j+1}} \\[4pt]
\mathbf{r}_{0b}^{11} = \dfrac{\partial \mathbf{p}(u,v)}{\partial u \partial v}\bigg|_{u=u_i, v=v_{j+1}} = \dfrac{1}{a}\dfrac{1}{b}(u_{i+1}-u_i)(v_{j+1}-v_j) \sum_{i=e-3}^{e-1} \sum_{j=f-2}^{f} \mathbf{d}_{i,j} (B_{i,3})_{,u}\big|_{u=u_i} (B_{j,3})_{,v}\big|_{v=v_{j+1}} \\[4pt]
\mathbf{r}_{ab}^{00} = \mathbf{p}(u_{i+1}, v_{j+1}) = \sum_{i=e-2}^{e} \sum_{j=f-2}^{f} \mathbf{d}_{i,j} B_{i,3}(u_{i+1}) B_{j,3}(v_{j+1}) \\[4pt]
\mathbf{r}_{ab}^{10} = \dfrac{\partial \mathbf{p}(u,v)}{\partial u}\bigg|_{u=u_{i+1}, v=v_{j+1}} = \dfrac{1}{a}(u_{i+1}-u_i) \sum_{i=e-2}^{e} \sum_{j=f-2}^{f} \mathbf{d}_{i,j} (B_{i,3})_{,u}\big|_{u=u_{i+1}} B_{j,3}(v_{j+1}) \\[4pt]
\mathbf{r}_{ab}^{01} = \dfrac{\partial \mathbf{p}(u,v)}{\partial v}\bigg|_{u=u_{i+1}, v=v_{j+1}} = \dfrac{1}{b}(v_{j+1}-v_j) \sum_{i=e-2}^{e} \sum_{j=f-2}^{f} \mathbf{d}_{i,j} B_{i,3}(u_{i+1}) (B_{j,3})_{,v}\big|_{v=v_{j+1}} \\[4pt]
\mathbf{r}_{ab}^{11} = \dfrac{\partial \mathbf{p}(u,v)}{\partial u \partial v}\bigg|_{u=u_{i+1}, v=v_{j+1}} = \dfrac{1}{a}\dfrac{1}{b}(u_{i+1}-u_i)(v_{j+1}-v_j) \sum_{i=e-2}^{e} \sum_{j=f-2}^{f} \mathbf{d}_{i,j} (B_{i,3})_{,u}\big|_{u=u_{i+1}} (B_{j,3})_{,v}\big|_{v=v_{j+1}}
\end{cases}
$$

(52)

in which, each basis function and corresponding derivative can be non-recursively represented in terms of Eq.(38). Eq.(53) shows these non-recursive basis functions and points one special letter $\theta$ and $\varphi$ with subscript to them for concise representation of Appendix B.

$$\begin{cases}
\theta_3 = B_{i-3,3}(\lambda_\omega) = \dfrac{H_{1,0}^2}{H_{1,-2}H_{1,-1}} \\[6pt]
\theta_2 = B_{i-2,3}(\lambda_\omega) = \dfrac{H_{0,-2}H_{1,0}}{H_{1,-2}H_{1,-1}} + \dfrac{H_{0,-1}H_{2,0}}{H_{2,-1}H_{1,-1}} \\[6pt]
\theta_1 = B_{i-1,3}(\lambda_\omega) = \dfrac{H_{0,-1}^2}{H_{2,-1}H_{1,-1}} \\[6pt]
\theta_{d3} = B'_{i-3,3}(\lambda_\omega) = \dfrac{-3H_{1,0}}{H_{1,-2}H_{1,-1}} \\[6pt]
\theta_{d2} = B'_{i-2,3}(\lambda_\omega) = \dfrac{H_{1,0}-2H_{0,-2}}{H_{1,-2}H_{1,-1}} + \dfrac{H_{-1,0}+2H_{2,0}}{H_{2,-1}H_{1,-1}} \\[6pt]
\theta_{d1} = B'_{i-1,3}(\lambda_\omega) = \dfrac{3H_{0,-1}}{H_{2,-1}H_{1,-1}} \\[6pt]
\varphi_2 = B_{i-2,3}(\lambda_{\omega+1}) = \dfrac{H_{2,1}^2}{H_{2,-1}H_{2,0}} \\[6pt]
\varphi_1 = B_{i-1,3}(\lambda_{\omega+1}) = \dfrac{H_{1,-1}H_{2,1}}{H_{2,-1}H_{2,0}} + \dfrac{H_{3,1}H_{1,0}}{H_{3,0}H_{2,0}} \\[6pt]
\varphi_0 = B_{i-0,3}(\lambda_{\omega+1}) = \dfrac{H_{1,0}^2}{H_{3,0}H_{2,0}} \\[6pt]
\varphi_{d2} = B'_{i-2,3}(\lambda_{\omega+1}) = \dfrac{-3H_{2,1}}{H_{2,-1}H_{2,0}} \\[6pt]
\varphi_{d1} = B'_{i-1,3}(\lambda_{\omega+1}) = \dfrac{H_{2,1}-2H_{1,-1}}{H_{2,-1}H_{2,0}} + \dfrac{2H_{3,1}-H_{1,0}}{H_{3,0}H_{2,0}} \\[6pt]
\varphi_{d0} = B'_{i-0,3}(\lambda_{\omega+1}) = \dfrac{3H_{1,0}}{H_{3,0}H_{2,0}}
\end{cases} \quad (53)$$

where, $B'$ represents the derivation of $B$. The definition of $H$ refers to Eq.(38). $\omega, \theta$ and $\varphi$ are defined as

$$\begin{cases} \omega = i, \theta = U^f, \varphi = U^l & \text{if } \lambda = u \\ \omega = j, \theta = V^f, \varphi = V^l & \text{if } \lambda = v \end{cases} \quad (54)$$

# Appendix B

$$\boldsymbol{\psi}_{11} = \begin{bmatrix} U_3^f V_3^f \mathbf{I} & U_3^f V_2^f \mathbf{I} & U_3^f V_1^f \mathbf{I} & \mathbf{0} \\ \upsilon U_{d3}^f V_3^f \mathbf{I} & \upsilon U_{d3}^f V_2^f \mathbf{I} & \upsilon U_{d3}^f V_1^f \mathbf{I} & \mathbf{0} \\ \zeta U_3^f V_{d3}^f \mathbf{I} & \zeta U_3^f V_{d2}^f \mathbf{I} & \zeta U_3^f V_{d2}^f \mathbf{I} & \mathbf{0} \\ \upsilon\zeta U_{d3}^f V_{d3}^f \mathbf{I} & \upsilon\zeta U_{d3}^f V_{d2}^f \mathbf{I} & \upsilon\zeta U_{d3}^f V_{d1}^f \mathbf{I} & \mathbf{0} \end{bmatrix} \quad (55)$$

$$\boldsymbol{\psi}_{12} = \begin{bmatrix} U_2^f V_3^f \mathbf{I} & U_2^f V_2^f \mathbf{I} & U_2^f V_1^f \mathbf{I} & \mathbf{0} \\ \upsilon U_{d2}^f V_3^f \mathbf{I} & \upsilon U_{d2}^f V_2^f \mathbf{I} & \upsilon U_{d2}^f V_1^f \mathbf{I} & \mathbf{0} \\ \zeta U_2^f V_{d3}^f \mathbf{I} & \zeta U_2^f V_{d2}^f \mathbf{I} & \zeta U_2^f V_{d2}^f \mathbf{I} & \mathbf{0} \\ \upsilon\zeta U_{d2}^f V_{d3}^f \mathbf{I} & \upsilon\zeta U_{d2}^f V_{d2}^f \mathbf{I} & \upsilon\zeta U_{d2}^f V_{d1}^f \mathbf{I} & \mathbf{0} \end{bmatrix} \quad (56)$$

$$\boldsymbol{\psi}_{13} = \begin{bmatrix} U_1^f V_3^f \mathbf{I} & U_1^f V_2^f \mathbf{I} & U_1^f V_1^f \mathbf{I} & \mathbf{0} \\ \upsilon U_{d1}^f V_3^f \mathbf{I} & \upsilon U_{d1}^f V_2^f \mathbf{I} & \upsilon U_{d1}^f V_1^f \mathbf{I} & \mathbf{0} \\ \zeta U_1^f V_{d3}^f \mathbf{I} & \zeta U_1^f V_{d2}^f \mathbf{I} & \zeta U_1^f V_{d1}^f \mathbf{I} & \mathbf{0} \\ \upsilon\zeta U_{d1}^f V_{d3}^f \mathbf{I} & \upsilon\zeta U_{d1}^f V_{d2}^f \mathbf{I} & \upsilon\zeta U_{d1}^f V_{d1}^f \mathbf{I} & \mathbf{0} \end{bmatrix} \quad (57)$$

$$\boldsymbol{\psi}_{14} = \begin{bmatrix} \mathbf{0} \end{bmatrix} \quad (58)$$

$$\boldsymbol{\psi}_{21} = \begin{bmatrix} \mathbf{0} \end{bmatrix} \quad (59)$$

$$\boldsymbol{\psi}_{22} = \begin{bmatrix} U_2^l V_3^f \mathbf{I} & U_2^l V_2^f \mathbf{I} & U_2^l V_1^f \mathbf{I} & \mathbf{0} \\ \upsilon U_{d2}^l V_3^f \mathbf{I} & \upsilon U_{d2}^l V_2^f \mathbf{I} & \upsilon U_{d2}^l V_1^f \mathbf{I} & \mathbf{0} \\ \zeta U_2^l V_{d3}^f \mathbf{I} & \zeta U_2^l V_{d2}^f \mathbf{I} & \zeta U_2^l V_{d2}^f \mathbf{I} & \mathbf{0} \\ \upsilon\zeta U_{d2}^l V_{d3}^f \mathbf{I} & \upsilon\zeta U_{d2}^l V_{d2}^f \mathbf{I} & \upsilon\zeta U_{d2}^l V_{d1}^f \mathbf{I} & \mathbf{0} \end{bmatrix} \quad (60)$$

$$\boldsymbol{\psi}_{23} = \begin{bmatrix} U_1^l V_3^f \mathbf{I} & U_1^l V_2^f \mathbf{I} & U_1^l V_1^f \mathbf{I} & \mathbf{0} \\ \upsilon U_{d1}^l V_3^f \mathbf{I} & \upsilon U_{d1}^l V_2^f \mathbf{I} & \upsilon U_{d1}^l V_1^f \mathbf{I} & \mathbf{0} \\ \zeta U_1^l V_{d3}^f \mathbf{I} & \zeta U_1^l V_{d2}^f \mathbf{I} & \zeta U_1^l V_{d1}^f \mathbf{I} & \mathbf{0} \\ \upsilon\zeta U_{d1}^l V_{d3}^f \mathbf{I} & \upsilon\zeta U_{d1}^l V_{d2}^f \mathbf{I} & \upsilon\zeta U_{d1}^l V_{d1}^f \mathbf{I} & \mathbf{0} \end{bmatrix} \quad (61)$$

$$\boldsymbol{\psi}_{24} = \begin{bmatrix} U_0^l V_3^f \mathbf{I} & U_0^l V_2^f \mathbf{I} & U_0^l V_1^f \mathbf{I} & \mathbf{0} \\ \upsilon U_{d0}^l V_3^f \mathbf{I} & \upsilon U_{d0}^l V_2^f \mathbf{I} & \upsilon U_{d0}^l V_1^f \mathbf{I} & \mathbf{0} \\ \zeta U_0^l V_{d3}^f \mathbf{I} & \zeta U_0^l V_{d2}^f \mathbf{I} & \zeta U_0^l V_{d1}^f \mathbf{I} & \mathbf{0} \\ \upsilon\zeta U_{d0}^l V_{d3}^f \mathbf{I} & \upsilon\zeta U_{d0}^l V_{d2}^f \mathbf{I} & \upsilon\zeta U_{d0}^l V_{d1}^f \mathbf{I} & \mathbf{0} \end{bmatrix} \quad (62)$$

$$\boldsymbol{\psi}_{31} = \begin{bmatrix} \mathbf{0} & U_3^f V_2^l \mathbf{I} & U_3^f V_1^l \mathbf{I} & U_3^f V_0^l \mathbf{I} \\ \mathbf{0} & \upsilon U_{d3}^f V_2^l \mathbf{I} & \upsilon U_{d3}^f V_1^l \mathbf{I} & \upsilon U_{d3}^f V_0^l \mathbf{I} \\ \mathbf{0} & \zeta U_3^f V_{d2}^l \mathbf{I} & \zeta U_3^f V_{d1}^l \mathbf{I} & \zeta U_3^f V_{d0}^l \mathbf{I} \\ \mathbf{0} & \upsilon\zeta U_{d3}^f V_{d2}^l \mathbf{I} & \upsilon\zeta U_{d3}^f V_{d1}^l \mathbf{I} & \upsilon\zeta U_{d3}^f V_{d1}^l \mathbf{I} \end{bmatrix} \quad (63)$$

$$\boldsymbol{\psi}_{32} = \begin{bmatrix} \mathbf{0} & U_2^f V_2^l \mathbf{I} & U_2^f V_1^l \mathbf{I} & U_2^f V_0^l \mathbf{I} \\ \mathbf{0} & \upsilon U_{d2}^f V_2^l \mathbf{I} & \upsilon U_{d2}^f V_1^l \mathbf{I} & \upsilon U_{d2}^f V_0^l \mathbf{I} \\ \mathbf{0} & \zeta U_2^f V_{d2}^l \mathbf{I} & \zeta U_2^f V_{d1}^l \mathbf{I} & \zeta U_2^f V_{d0}^l \mathbf{I} \\ \mathbf{0} & \upsilon\zeta U_{d2}^f V_{d2}^l \mathbf{I} & \upsilon\zeta U_{d2}^f V_{d1}^l \mathbf{I} & \upsilon\zeta U_{d2}^f V_{d0}^l \mathbf{I} \end{bmatrix} \quad (64)$$

$$\boldsymbol{\psi}_{33} = \begin{bmatrix} \mathbf{0} & U_1^f V_2^l \mathbf{I} & U_1^f V_1^l \mathbf{I} & U_1^f V_0^l \mathbf{I} \\ \mathbf{0} & \upsilon U_{d1}^f V_2^l \mathbf{I} & \upsilon U_{d1}^f V_1^l \mathbf{I} & \upsilon U_{d1}^f V_0^l \mathbf{I} \\ \mathbf{0} & \zeta U_1^f V_{d2}^l \mathbf{I} & \zeta U_1^f V_{d1}^l \mathbf{I} & \zeta U_1^f V_{d0}^l \mathbf{I} \\ \mathbf{0} & \upsilon\zeta U_{d1}^f V_{d2}^l \mathbf{I} & \upsilon\zeta U_{d1}^f V_{d1}^l \mathbf{I} & \upsilon\zeta U_{d1}^f V_{d0}^l \mathbf{I} \end{bmatrix} \quad (65)$$

$$\boldsymbol{\psi}_{34} = \begin{bmatrix} \mathbf{0} \end{bmatrix} \quad (66)$$

$$\boldsymbol{\psi}_{41} = \begin{bmatrix} \mathbf{0} \end{bmatrix} \quad (67)$$

$$\boldsymbol{\psi}_{42} = \begin{bmatrix} \mathbf{0} & U_2^l V_2^l \mathbf{I} & U_2^l V_1^l \mathbf{I} & U_2^l V_0^l \mathbf{I} \\ \mathbf{0} & \upsilon U_{d2}^l V_2^l \mathbf{I} & \upsilon U_{d2}^l V_1^l \mathbf{I} & \upsilon U_{d2}^l V_0^l \mathbf{I} \\ \mathbf{0} & \zeta U_2^l V_{d2}^l \mathbf{I} & \zeta U_2^l V_{d1}^l \mathbf{I} & \zeta U_2^l V_{d0}^l \mathbf{I} \\ \mathbf{0} & \upsilon\zeta U_{d2}^l V_{d2}^l \mathbf{I} & \upsilon\zeta U_{d2}^l V_{d1}^l \mathbf{I} & \upsilon\zeta U_{d2}^l V_{d0}^l \mathbf{I} \end{bmatrix} \quad (68)$$

$$\boldsymbol{\psi}_{43} = \begin{bmatrix} \mathbf{0} & U_1^l V_2^l \mathbf{I} & U_1^l V_1^l \mathbf{I} & U_1^l V_0^l \mathbf{I} \\ \mathbf{0} & \upsilon U_{d1}^l V_2^l \mathbf{I} & \upsilon U_{d1}^l V_1^l \mathbf{I} & \upsilon U_{d1}^l V_0^l \mathbf{I} \\ \mathbf{0} & \zeta U_1^l V_{d2}^l \mathbf{I} & \zeta U_1^l V_{d1}^l \mathbf{I} & \zeta U_1^l V_{d0}^l \mathbf{I} \\ \mathbf{0} & \upsilon\zeta U_{d1}^l V_{d2}^l \mathbf{I} & \upsilon\zeta U_{d1}^l V_{d1}^l \mathbf{I} & \upsilon\zeta U_{d1}^l V_{d0}^l \mathbf{I} \end{bmatrix} \quad (69)$$

$$\boldsymbol{\psi}_{44} = \begin{bmatrix} \mathbf{0} & U_0^l V_2^l \mathbf{I} & U_0^l V_1^l \mathbf{I} & U_0^l V_0^l \mathbf{I} \\ \mathbf{0} & \upsilon U_{d0}^l V_2^l \mathbf{I} & \upsilon U_{d0}^l V_1^l \mathbf{I} & \upsilon U_{d0}^l V_0^l \mathbf{I} \\ \mathbf{0} & \zeta U_0^l V_{d2}^l \mathbf{I} & \zeta U_0^l V_{d1}^l \mathbf{I} & \zeta U_0^l V_{d0}^l \mathbf{I} \\ \mathbf{0} & \upsilon\zeta U_{d0}^l V_{d2}^l \mathbf{I} & \upsilon\zeta U_{d0}^l V_{d1}^l \mathbf{I} & \upsilon\zeta U_{d0}^l V_{d0}^l \mathbf{I} \end{bmatrix} \quad (70)$$

In Eq. (55)~ Eq.(70), $\upsilon = \frac{1}{a}(u_{i+1} - u_i)$ ; $\zeta = \frac{1}{b}(v_{j+1} - v_j)$.